\documentclass[twocolumn,preprintnumbers,amsmath,amssymb]{revtex4}

\usepackage{graphicx}
\usepackage{dcolumn}
\usepackage{bm}

\begin{document}

\title{Nonlinear effects in charge stabilized colloidal suspensions}

\author{T. Kreer, J. Horbach, A. Chatterji}

\address{
Institut f\"ur Physik, Johannes Gutenberg-Universit\"at, 
55099 Mainz, Germany\\
}

\date{\today}

\begin{abstract} 
Molecular Dynamics simulations are used to study the effective
interactions in charged stabilized colloidal suspensions. For not
too high macroion charges and sufficiently large screening, the
concept of the potential of mean force is known to work well. In the
present work, we focus on highly charged macroions in the limit of low
salt concentrations. Within this regime, nonlinear corrections to the
celebrated DLVO theory [B. Derjaguin and L. Landau, Acta Physicochem. USSR
{\bf 14}, 633 (1941); E.J.W. Verwey and J.T.G. Overbeck, {\em Theory
of the Stability of Lyotropic Colloids} (Elsevier, Amsterdam, 1948)]
have to be considered. For non--bulklike systems, such as isolated
pairs or triples of macroions, we show, that nonlinear effects can
become relevant, which cannot be described by the charge renormalization
concept [S. Alexander et al., J. Chem. Phys. {\bf 80}, 5776 (1984)]. For
an isolated pair of macroions, we find an almost perfect qualitative
agreement between our simulation data and the primitive model. However,
on a quantitative level, neither Debye-H\"uckel theory nor the charge
renormalization concept can be confirmed in detail. This seems mainly to
be related to the fact, that for small ion concentrations, microionic
layers can strongly overlap, whereas, simultaneously, excluded volume
effects are less important. In the case of isolated triples, where we
compare between coaxial and triangular geometries, we find attractive
corrections to pairwise additivity in the limit of small macroion
separations and salt concentrations. These triplet interactions arise if
all three microionic layers around the macroions exhibit a significant
overlap.  In contrast to the case of two isolated colloids, the charge
distribution around a macroion in a triple is found to be anisotropic.
\end{abstract}

\maketitle

\section{Introduction}
\label{sec:intro}
In order to simplify the description of colloidal systems, one often
tries to determine effective interactions between the colloidal
particles, thus integrating out the solvent's degrees of freedom
\cite{belloni00}. This is not a trivial task, because in general the
effective interactions depend on the thermodynamic state of
the system, and one is often confronted with the problem of thermodynamic
inconsistencies~\cite{louis02}.  A problem that is of particular interest
is that of effective interactions between charged colloids (macroions)
in an electrolyte solution.  On a mean--field level, such a system can
be described by the Poisson--Boltzmann (PB) equation \cite{russel}. The
linearized version of this equation is the basis of the DLVO theory for
charged colloids~\cite{dlvo}, and we will refer to it in the following
as the Debye--H\"uckel (DH) limit of the PB equation \cite{debye23}.
In the DH limit, the problem can be solved analytically and yields a
screened Coulomb potential for the effective interactions between the
macroions.  The characteristic range of this potential is given by the
Debye length $\kappa^{-1}$, which is controled by solvent properties
such as the salt concentration.

The physical picture of the DH description is rather appealing: Due to the
charge of a macroion, a layer of thickness $\kappa^{-1}$ is formed around
it, consisting of oppositely charged microions (counterions), leading to
a screening of the bare Coulomb interaction.  Although the DH description
is only valid for weakly charged macroions and if correlation effects
between the microions in the electrolyte solution can be neglected,
it is tempting to characterize also the effective interactions between
highly charged macroions by a potential of screened Coulomb form. Indeed,
this is the idea of the famous concept of charge renormalization that
has been put forward by Alexander {\it et al.}~\cite{alex}. It is based
on the observation that in the framework of the so--called cell model
(see below) the numerical solution of the nonlinear PB equation can
be fitted farther away from the macroions' boundaries by a screened
Coulomb potential with a renormalized charge $Z_{\rm eff}<Z$ (with
$Z$ the bare charge of a macroion). Trizac {\it et al.}~\cite{trizac}
recently extended the numerical recipe of Alexander {\it et al.}~by
providing an analytical scheme to calculate $Z_{\rm eff}$ as well as
the effective screening length and the effective salt concentration.

As already mentioned, DH theory is based on the linearized PB equation,
which implies pairwise additivity of interaction energies.  On the other
hand, for highly charged macroions, nonlinearities imply the occurrence
of many--body interactions and thus pairwise additivity does not hold. The
simplest system, in which many--body effects could be expected, consists
of three isolated macroions in an electrolyte solution. Indeed, such
a system has been studied in a recent experiment using scanned optical
line tweezers \cite{brunner,dobnikar}. In this work, charge--stabilized
silica particles with a diameter of about 1\,$\mu$m suspended in water
were considered. It was possible to measure three--body interactions
directly, and it was found that, in agreement with numerical solutions
of the nonlinear PB equation \cite{belloni00,denton04}, three--body
contributions to the total interaction energy are attractive.

Also Molecular Dynamics (MD) computer simulations have been used to
investigate systems of ``isolated'' macroion pairs and triples
\cite{allah1,lowen,allah2,tehver,shapran05}. In these studies, charged
colloids were investigated in the framework of the so--called primitive
model. In this model, a system of macroions, counterions and salt ions is
considered without explicitly taking into account the uncharged part of
the solvent.  Based on the primitive model, Allahyarov and L\"owen found
that DH theory works well for a system of two macroions \cite{allah1},
and, in agreement with experiment \cite{brunner,dobnikar} and PB theory
\cite{russ}, that three--body contributions are attractive in the case
of three macroions \cite{lowen}.  These authors also studied a system
of two macroions, in which uncharged solvent particles were added to the
electrolyte solution \cite{allah2}.  An interesting finding of this work
was that the neutral solvent leads to a renormalized charge,
which is smaller than the bare charge of the macroions, similar to the
concept proposed by Alexander {\it et al.}~\cite{alex}. In a different
simulation study by Tehver {\it et al.} \cite{tehver}, the counterions
were introduced via density distributions in the framework of a density
functional theory. Surprisingly, in the case of three macroions, no
evidence for many--body forces was found, and the forces could be well
described by DH theory.

In this work, MD simulations are presented that tie in with the previous
simulation studies. We consider systems of two and three highly charged
macroions in a primitive model solvent. In the two--particle case, we
check to what extent DH theory describes the effective interactions;
thereby, the effect of nonlinearities is quantified. This is done for
different amounts of added salt, focussing on the limit of small salt
concentrations.  In a second step, we address the influence of nonlinear
effects on three-body interactions. In order to study these effects, a
triple of macroions is considered in two different geometries by
placing the macroions on an equililateral triangle or along a straight
line.  We check on whether the concept of charge renormalization can
also be applied to isolated pairs or triples of particles. Furthermore,
we ask for the validity of the mean--field description and how effective
interactions develop from the two--particle case to the bulk. Our major
concern is the influence of nonlinearity, which can be seen for high
macroion charges and low salt concentrations. We especially consider cases
of overlapping or interacting Debye-layers in the case of non--bulklike
macroion configurations.

Our paper is organized as follows: After briefly discussing some
results of DH theory and the concept of charge renormalization, we
give an overview of the simulation details.  In Sec.~\ref{sec:res}A we
present our results for systems that consist of a pair of macroions,
and in Sec.~\ref{sec:res}B systems with macroion triples are considered.
Finally, we discuss the results and draw some conclusions.

\section{DH potential, PB equation, and the concept of charge renormalization}
In this section, we consider charged spherical macroions of diameter
$\sigma$ and positive charge $Ze$ (here, $e$ is the elementary charge
and $Z$ the valency). They are immersed into a polar, structureless
medium with dielectric constant $\epsilon$. This medium is characterized
by the Bjerrum length $\lambda_{\rm B}= e^2/(4 \pi \epsilon k_{\rm B} T)$,
i.e.~the distance at which the electrostatic energy between two point
charges equals the thermal energy $k_{\rm B} T$.

In the DH limit, the interaction potential between two macroions,
separated by a distance $r$, is given by a screened Coulomb (Yukawa)
potential \cite{russel},
\begin{equation}
  \label{yukawa}
  u(r)=k_{\rm B} T \lambda_{\rm B} 
       \left[ \frac{Z \exp(\kappa\sigma/2)}{1+\kappa\sigma/2}\right]^2
       \frac{\exp(-\kappa r)}{r},
\end{equation}
where
\begin{equation}
  \label{kappa}
  \kappa=\sqrt{4\pi\lambda_{\rm B}(2 n_{\rm s}+ Z n_{\rm c})/V} \quad 
\end{equation}
is the screening parameter, $n_{\rm c}$ represents the number of
macroions, and $n_{\rm s}$ is the number of added salt ion pairs. 
In Eq.~(\ref{kappa}), it is assumed that the electrolyte
solution is formed by monovalent microions in a system of total volume
$V$.  The microions consist of $Zn_{\rm c}$ negatively charged counterions
that neutralize the charge of the macroions and $2 n_{\rm s}$ salt ions,
consisting half--and--half of counterions and oppositely charged coions.
The inverse of the screening parameter, the so--called Debye length
$R_{\rm D}=1/\kappa$, ``measures'' the thickness of the neutralizing
counterion layer around the macroions. Equation (\ref{kappa}) shows that
$R_{\rm D}$ can be varied by changing the properties of the solvent,
in particular the salt concentration.

Alexander {\it et al.}~\cite{alex} have demonstrated that many charged
colloidal systems with highly charged macroions can be described to some
extent by a Yukawa potential of the form of Eq.~(\ref{yukawa}), although
the DH limit is restricted to particles with small charge.  This is due to
the fact that highly charged colloids have a strong tendency to form ordered
structures at relatively low densities, i.e.~at densities where the mean
distance between neighboring macroions is much larger than their size. In
such systems, each macroion has a very similar environment of microions,
and thus a reasonable approximation is to reduce the problem of computing
the effective many--particle interactions between macroions to that of
determining a mean--field potential of one particle in its Wigner--Seitz
(WS) cell surrounded by a reservoir of salt ions \cite{alex}. For spherical 
macroions, the WS cell is approximated by a sphere of radius $R$. Then, one 
considers the nonlinear PB equation for the single particle with appropriate
boundary conditions \cite{alex},
\begin{eqnarray}
 \nabla^2 u(r) = & & \hspace*{-0.5cm} \frac{e \rho_{\rm s}}{4 \pi \epsilon} 
  \left[ \exp \left( \frac{e u(r)}{k_{\rm B} T} \right) +
         \exp \left( - \frac{e u(r)}{k_{\rm B} T} \right) \right] \\
  & & \quad \quad \quad \quad \quad \quad \quad \quad
(\sigma/2 < r < R \label{pb1}) \nonumber \\
 \vec{n} \cdot \nabla u(r) & = & \frac{Ze}{\pi \epsilon \sigma^2} 
  \quad \quad (r=\sigma/2) \label{pb2} \\
 \vec{n} \cdot \nabla u(r) & = & 0 \quad \quad (r=R) \label{pb3}   
\end{eqnarray}
with $\vec{n}$ the normal vector pointing outwards the sphere's surface
and $\rho_{\rm s}$ the salt concentration in the reservoir.  Equations
(\ref{pb1}) and (\ref{pb3}) are solved numerically. Then, one assumes that at
the cell boundary, i.e.~far away from the surface of the particle, the
solution $u(r)$ of the PB equation can be approximated by an effective
Yukawa potential,
\begin{equation}
  \label{effyukawa}
  u_{\rm eff}(r)=k_{\rm B} T \lambda_{\rm B} 
       \left[ \frac{Z_{\rm eff} 
       \exp(\kappa_{\rm eff}\sigma/2)}{1+\kappa_{\rm eff}\sigma/2}\right]^2
       \frac{\exp(-\kappa_{\rm eff} r)}{r}.
\end{equation}
The parameters $Z_{\rm eff}$ and $\kappa_{\rm eff}$ can be fixed by
matching the effective potential $u_{\rm eff}$ at the cell boundary
($r=R$) with that of the solution of the nonlinear PB equation.
In the original paper by Alexander {\it et al.}~\cite{alex}, this
is achieved by the following recipe: The screening parameter
$\kappa_{\rm eff}$ is determined by the microion densities
$n_{\pm}^{R} = \rho_0 \exp\left(\pm \frac{e u(r)}{k_{\rm B} T} \right)$ at
the WS cell boundary via $\kappa_{\rm eff}^2= 4 \pi \lambda_{\rm B}
(n_{+}^{R}+n_{-}^{R})$.  The effective charge $Z_{\rm eff}$ is fixed as
follows: First, Eqs.~(\ref{pb1}) and (\ref{pb3}) are linearized at $r=R$. For
the linearized equation a solution is determined such that the linear
and the nonlinear solution match up to the second derivative at the cell
boundary. Finally, $Z_{\rm eff}$ is calculated from the integral over
the charge density associated with the linear solution.

A more elegant recipe to obtain $Z_{\rm eff}$ and $\kappa_{\rm eff}$
has recently been proposed by Trizac {\it et al.}~\cite{trizac}. They
show that the full numerical solution of the nonlinear PB equation is
not needed to estimate the latter parameters. Instead, only the solution
$u_R$ at the cell boundary is required. Thus, only linearized equations
have to be solved, and this can be done analytically.  For $Z_{\rm eff}$,
Trizac {\it et al.}~\cite{trizac} find
\begin{eqnarray}
\label{eqzeff}
Z_{\rm eff} & = &
\frac{\gamma_0}{\kappa_{\rm eff}\lambda_{\rm B}}
[(\kappa_{\rm eff}^2 \sigma R/2-1)
\sinh(\kappa_{\rm eff}(R-\sigma/2)\\
& + &\nonumber 
\kappa_{\rm eff}(R-\sigma/2)\cosh(\kappa_{\rm eff}(R-\sigma/2)],
\end{eqnarray}
where $\gamma_0={\rm tanh}(u_R)$. Equation~(\ref{eqzeff}) implies 
$Z_{\rm eff}\le Z$, where
the effective charge (also called ``renormalized charge'') can be an
order of magnitude smaller than the bare charge.

The effective screening parameter $\kappa_{\rm eff}$ is related
to the effective salt concentration and $Z_{\rm eff}$ via \cite{trizac}
\begin{equation}
\label{eqnseff}
n_{\rm s}^{\rm eff}/V=
\frac{\kappa_{\rm eff}^2}{8\pi\lambda_{\rm B}}(1-\gamma_0^2)
(1-\eta)-\frac{1}{2V}Z_{\rm eff}n_{\rm c}(1-\gamma_0), 
\end{equation}
where $n_{\rm c}$ represents the number of macroions per WS cell.
The physical interpretation of the latter equation is related to the
so-called Donnan effect. Since a macroion occupies a finite volume
inside its WS cell, the microions are partially expelled. Thus, the
salt concentration outside the WS cell can be higher than inside the
colloid compartement, or, in other words, there is an effective salt
concentration which is smaller than the actual one, i.e.~$n_{\rm s}^{\rm
eff}<n_{\rm s}$.

For dilute systems, where the Donnan effect should not be relevant,
i.e.~for $\gamma_0$,$\eta\rightarrow0$, the effective salt concentration
matches the actual one, and hence Eq.~(\ref{eqnseff}) can be rewritten as
\begin{equation}
\label{eq:alex}
\kappa_{\rm eff}^2=4\pi\lambda_{\rm B}(Z_{\rm eff}n_{\rm c}+2n_{\rm s})/V.  
\end{equation}
Thus, setting $n_{\rm s}^{\rm eff}=n_{\rm s}$, leads back to an
one--parameter problem.  It follows from Eq.~(\ref{eq:alex}) that
for monovalent microions, $Z_{\rm eff}=Z$ also implies $\kappa_{\rm
eff}=\kappa$.

We emphasize that the systems considered in the following do not
match with the assumptions made in Alexander's concept of charge
renormalization.  In this work, ``non--bulklike'' systems are considered,
for which the definition of a WS cell is not meaningful. Moreover,
whereas charge renormalization is applied to distances far away from the
surface of a macroion, we are interested in relatively small distances
between macroions and thus also in the potential of mean force close to
their surfaces.

\section{Details of the simulation}
\label{sec:model}
Using classical MD simulations, we study charged
colloidal suspensions in the framework of the so--called primitive model.
We consider systems of two or three positively charged macroions of
valency $Z \equiv Z_{\rm m}=255$ and monovalent microions of charge
$Z_{\rm ct}e=-1$ (counterions) and of charge $Z_{\rm co}e=+1$ (coions). The
interaction potential between an ion of type $\alpha$ and an ion of type
$\beta$ ($\alpha, \beta = {\rm m, ct, co}$), separated by a distance $r$
from each other, is given by
\begin{equation}
  u_{\alpha \beta} =
  \frac{Z_{\alpha} Z_{\beta} e^2}{4 \pi \epsilon r}
  + A_{\alpha \beta}
    \exp \left\{ - B_{\alpha \beta} (r - \sigma_{\alpha \beta} )
    /\sigma_{\alpha \beta} \right\},
  \label{eq_pot}
\end{equation}
where the dielectric constant is set to $\epsilon = 79 \epsilon_0$ (with
$\epsilon_0$ the vacuum dielectric constant), which corresponds to the
value for water at room temperature. The parameter $\sigma_{\alpha \beta}$
is the distance between two ions at contact, $\sigma_{\alpha \beta}=
R_{\alpha} + R_{\beta}$, where $R_{\alpha}$ is the radius of an ion of
type $\alpha$. In our simulations, we used $R_{\rm m}=10$\,nm and $R_{\rm
ct}=R_{\rm co}=0.01 R_{\rm m}$. The choice of the latter values guarantees
that depletion effects are not relevant. The exponential in Eq.~(\ref{eq_pot}) is
an approximation to a hard sphere interaction for two ions at contact.
For the parameters $A_{\alpha \beta}$ we chose $A_{\rm mm}=1.84$~eV,
$A_{\rm mct}=A_{\rm mco}=0.0556544$~eV, and $A_{\rm ctct}=A_{\rm ctco}=A_{\rm
coco}=0.0051$~eV. The parameters $B_{\alpha \beta}$ are all set to $3$.
The long--ranged Coulomb part of the potential and the forces were
computed by Ewald sums in which we chose $\alpha=0.05$ for the constant
and a cutoff wavenumber $k_{\rm c}=2\pi\sqrt{66}/L$ in the Fourier
part~\cite{allen}.  The linear dimension $L$ of the simulation box is 
$L=159.026$\,nm, using periodic boundary conditions. 

Since the potential, Eq.~(\ref{eq_pot}), is long--ranged, one has to
consider the possible emergence of finite--size effects. However, in our
simulations, the distance of a macroion to its next periodic image was
always larger than $7\,\sigma$ (with $\sigma\equiv \sigma_{\rm mm}$),
and, as discussed in the next section, at this distance the Coulomb
interaction is sufficiently screened.  Instead of using periodic boundary
conditions, an alternative approach would be to confine the system by
walls~\cite{allah1,lowen}. However, due to the interaction of the ions
with the walls, also in this case finite--size effects are relevant
(indeed in Refs.~\cite{allah1,lowen}, a correction term had to be
introduced to estimate the ``bulk'' effective force between macroions).

\begin{figure}[t]
\includegraphics[width=6.cm]{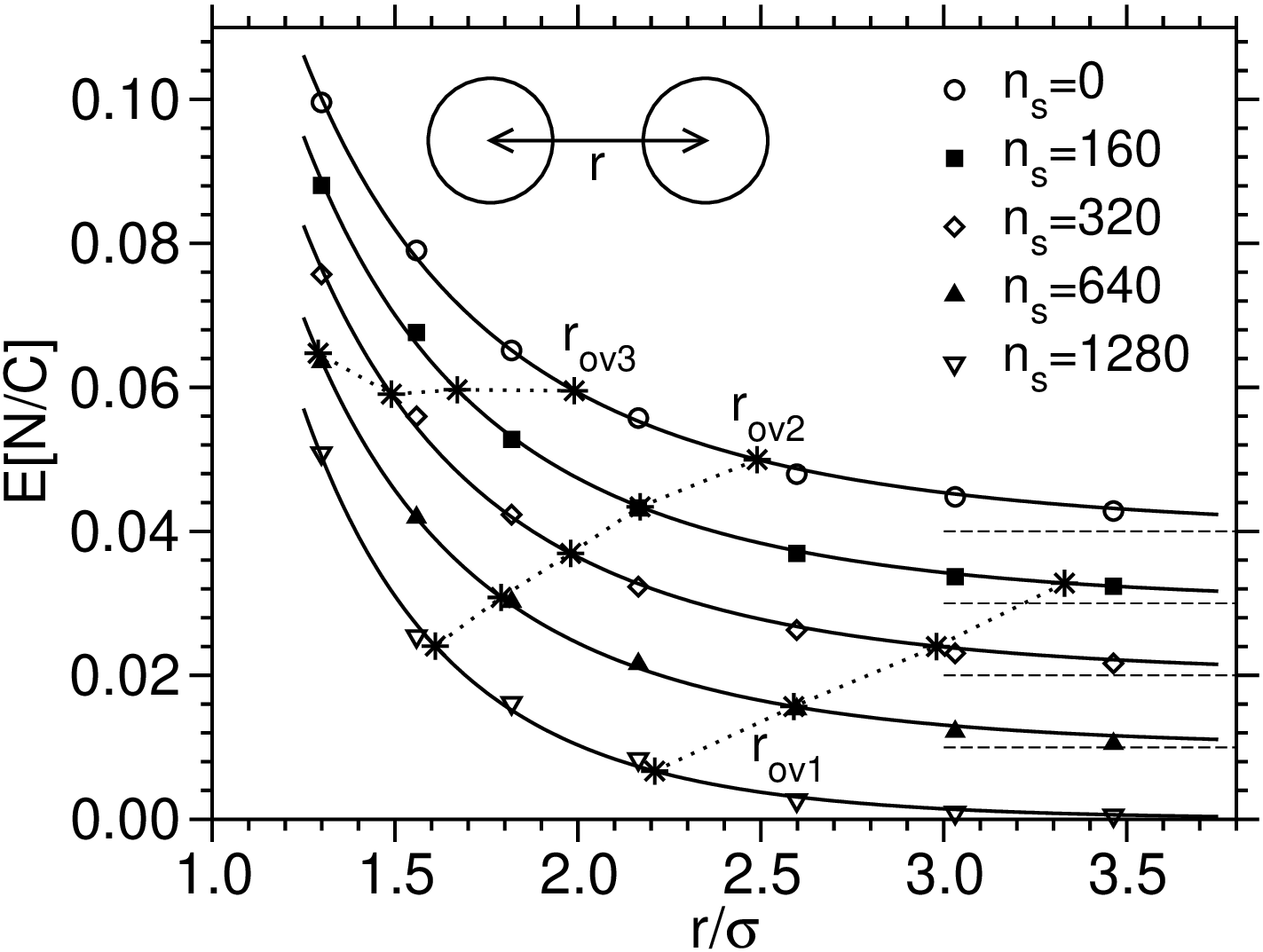}
\caption{
Electric field around macroion as a function of distance for indicated
salt concentrations. Solid lines are fits to Eq.~(\ref{effyukawa}), where
$Z_{\rm eff}$ and $\kappa_{\rm eff}$ are used as fit parameters. Data
is shifted such that the dotted lines represent an electric field of
$E=0$ for each value of $n_{\rm s}$. Stars indicate critical macroion
separations, as defined in text. Dashed lines are guides to the
eye. Statistical errors are smaller than twice the size of symbols.
\label{fig1}
}
\end{figure}
In order to determine the effective forces between macroions at a
distance $r$, the macroions are fixed by decoupling macroionic and
microionic time scales. This is achieved by assigning a mass to the
macroion which is $10^6 m_{\rm ct}$ (with $m_{\rm ct}=m_{\rm co}$
the mass of the microions).  All the simulations were done at the
temperature $T=298$\,K.  Thus, the Bjerrum length for our system is
$\lambda_{\rm B}\approx 0.71{\rm nm}$.  The number of added coions was
varied from $n_{\rm s}=0$ to $n_{\rm s}=1280$.  Being $n_{\rm c}'$ the
number of macroions in the system, charge neutrality requires $n_{\rm
tot}=Zn_{\rm c}'+2n_{\rm s}$ for the total number of microions.

The equations of motion were integrated using the velocity form of the
Verlet algorithm. The simulations were done at constant temperature. In
order to thermostat the system, it was coupled to a stochastic
heat bath \cite{allen}.  For a given set of parameters ($n_{\rm s}$,
macroion separation $r$), we examined at least three independent start
configurations.  For equilibration, runs of $10^5-10^6$ MD time steps
were done, followed by a similar number of steps to calculate time
averages. Depending on salt concentration, the time step varied from 
$\Delta t=1\cdot 10^{-4}\tau_{0}$ to $\Delta t=3\cdot 10^{-4}\tau_{0}$
[with $\tau_{0}\equiv R_{\rm ct}\sqrt{m_{\rm ct}/(k_{\rm B} T)}$].

\section{Results and Discussion}
\label{sec:res}
\subsection{Two macroions}
\begin{figure}[t]
\includegraphics[width=6.cm]{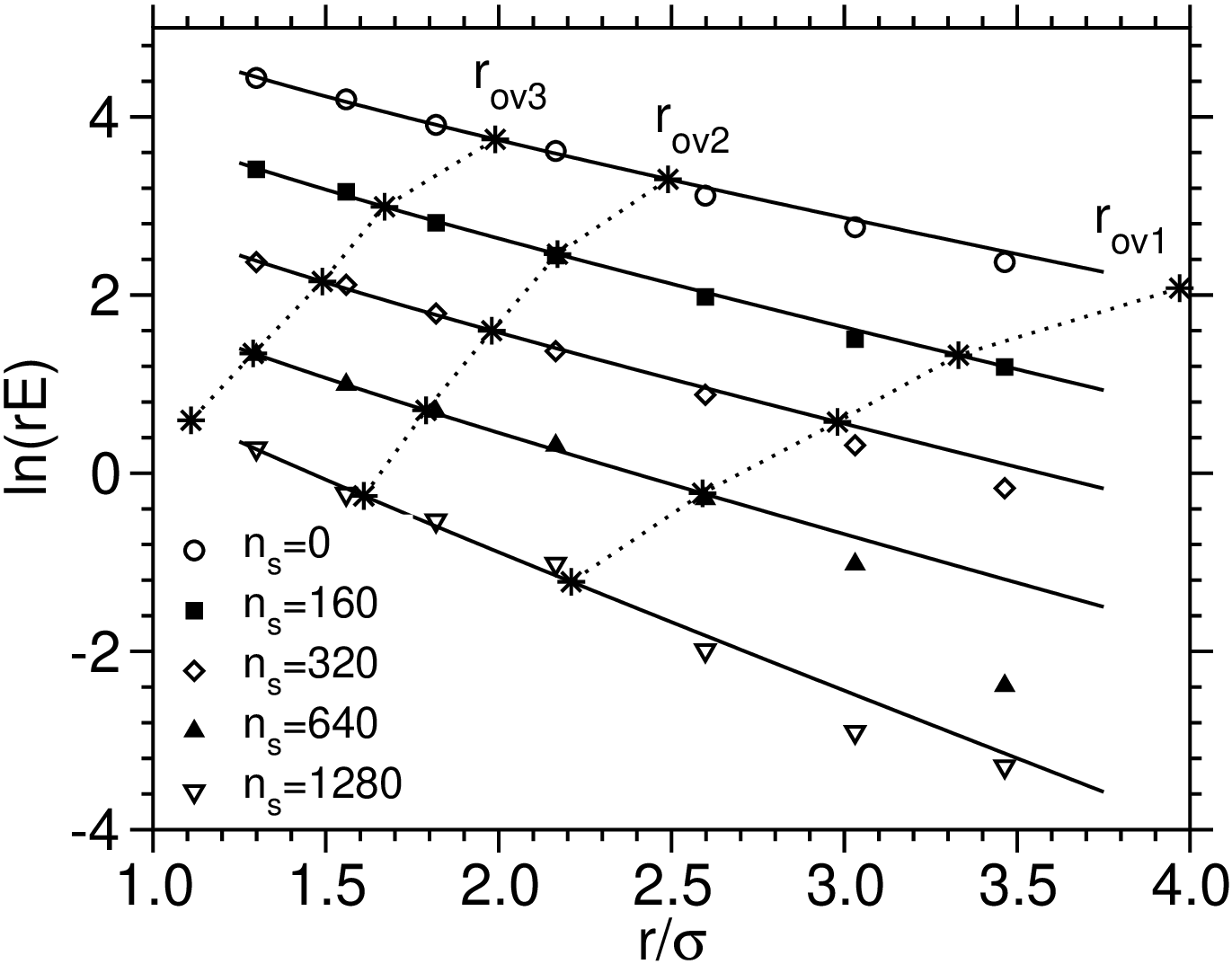}
\caption{
Logarithm of electric field times macroion separation versus $r$ for
various salt concentrations. Solid lines are fits to Eq.~(\ref{effyukawa})
using the same fitting parameters as in Fig.~\ref{fig1}.
Data for $n_{\rm s}=0$ is shifted by $4$, $n_{\rm s}=160$ by $3$,
$n_{\rm s}=320$ by $2$, and $n_{\rm s}=640$ by $1$, respectively. Stars
and dashed lines have same meaning as in Fig.~\ref{fig1}.
\label{fig2}
}
\end{figure}
The effective interaction between two macroions can be quantified
by the electric field $E(r)$ around a macroion, depending on the
separation $r$ between the macroion's centers. The field $E$ is given by
$E(r)=\frac{1}{Ze}F(r)$, where $F(r)=-\frac{\partial V(r)}{\partial r}$
is the total force on a macroion projected onto the line that connects
the macroion's centers. Figure \ref{fig1} shows $E(r)$ for different
salt concentrations, as indicated.  A comparison to the prediction
from DH theory, using $u_{\rm eff}(r)$ from Eq.~(\ref{effyukawa}), with
$Z_{\rm eff}$ and $\kappa_{\rm eff}$ as fitting parameters, reveals a
good agreement.  However, as we will show in the following, the numerical
values we obtain for $Z_{\rm eff}$ and $\kappa_{\rm eff}$ can neither
be described by the DH predictions nor by the charge renormalization
concept, as it results in Eqs.~(\ref{eqzeff}) and (\ref{eqnseff}).
Plotting our data on a logarithmic scale exposes deviations from the
DH form (Fig.~\ref{fig2}), which seem to increase with distance and
salt concentration.  We confirmed the absence of finite size effects
by repeating some of our simulation runs in a system of double volume,
finding our results to be consistent and not depending on the volume.

To analyze our data further, it is useful to quantify the extent to what
the Debye layers around the macroions overlap.  To this end, we use the
DH expression, Eq.~(\ref{kappa}), to estimate the inverse Debye layer
thickness. According to this definition of $\kappa$, Debye layers overlap,
if $\frac{\kappa\sigma}{2}(\frac{r}{\sigma}-1)<1$. For a given value of
$n_{\rm s}$ we thus define $r_{\rm ov}^{(1)}\equiv\sigma+2\kappa^{-1}$
as an upper critical macroion separation. For
$\kappa\sigma(\frac{r}{\sigma}-1)<1$, the macroions are (partially)
located within each other's Debye layers. For a fixed salt concentration,
we therefore define $r_{\rm ov}^{(2)}\equiv\sigma+\kappa^{-1}$.
Finally, we consider the limit, where a macroion's center of mass
is located within the Debye layer of the other macroion. This is the
case for $\kappa\sigma(\frac{r}{\sigma}-\frac{1}{2})<1$, or, $r_{\rm
ov}^{(3)}\equiv\sigma/2+\kappa^{-1}$. The radii $r_{\rm ov}^{(1,2,3)}$
are indicated in Figs.~\ref{fig1} and \ref{fig2} as stars (connected by
dashed lines as a guide to the eye).

As we see in Figs.~\ref{fig1} and \ref{fig2}, Debye layers overlap
for almost all parameter combinations considered.  Thus, we find a
Yukawa--like potential also in the region of strongly overlapping
Debye layers. The Yukawa--form persists even for macroion-separations
$r<r_{\rm ov}^{(3)}$, where the macroions are relatively close to
contact. These findings are in agreement with numerical solutions of
the nonlinear PB equation for a system of two macroions~\cite{russ}.
The significant deviations from the DH fits seem to occur for
non--overlapping Debye layers (see Fig.~\ref{fig2}). This might be related
to the low signal--to--noise ratio, which becomes worse for increasing
values of $\kappa r$.

\begin{figure}[t]
\includegraphics[width=8.cm]{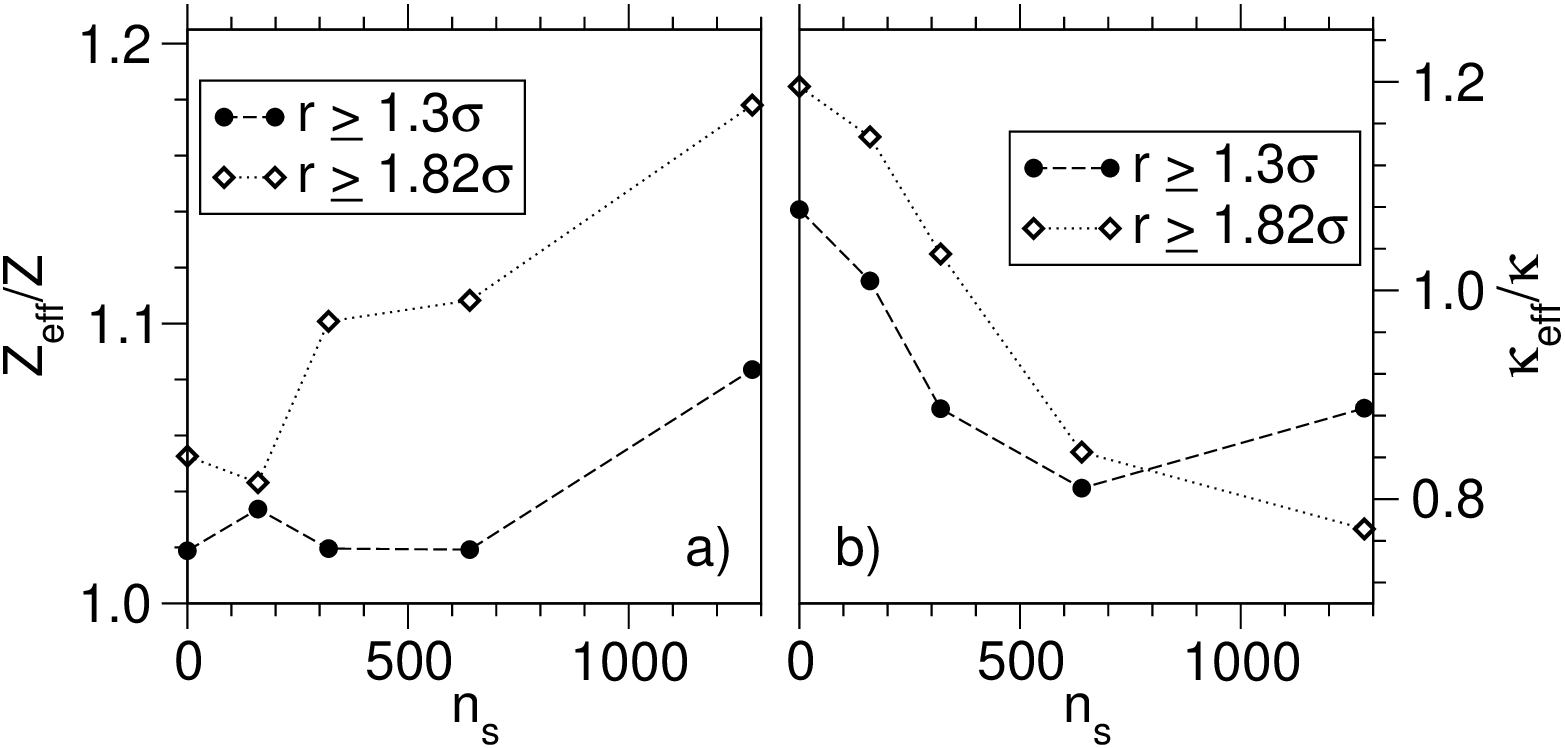}
\caption{
a) Effective charge $Z_{\rm eff}$ normalized by the bare charge $Z$ and b)
effective screening parameter $\kappa_{\rm eff}$ divided by $\kappa$
from DH theory [Eq.~(\ref{kappa})] as a function of the number of salt
ion pairs $n_{\rm s}$. The parameters $Z_{\rm eff}$ and $\kappa_{\rm
eff}$ result from fits shown in Fig.~\ref{fig1} and Fig.~\ref{fig2}
that include, as indicated, all data points with $r\ge1.3\sigma$ or only
those with $r\ge 1.82\sigma$.
\label{fig3}
}
\end{figure}
Fig.~\ref{fig3} shows the fitted values of $Z_{\rm eff}$ and $\kappa_{\rm
eff}$ as a function of salt ion pairs $n_{\rm s}$.  Here, we have
normalized $Z_{\rm eff}$ by the bare charge of the macroions and
$\kappa_{\rm eff}$ by the screening parameter $\kappa$ as predicted
by Eq.~(\ref{kappa}). The fit values in Fig.~\ref{fig3} are extracted
from two different types of fits. In addition to fits that include all
available data points, we performed also fits that are restricted to data
points with macroion separations of $r \ge 1.82\sigma > r_{\rm ov}^{(3)}$.
Thus, in the latter fits, we exclude distances for which the center of
a macroion penetrates into the Debye layer of the other one.

As one can infer from Fig.~\ref{fig3}, $\kappa_{\rm eff}/\kappa$
and $Z_{\rm eff}/Z$ deviate significantly from unity for all salt
concentrations considered (except for $n_{\rm s}\approx 250$ where
$\kappa_{\rm eff}/\kappa$ is close to one). Thus, for the systems under
consideration, DH theory does not correctly describe the effective
interactions. This indicates that a combination of nonlinear effects
and, possibly, microionic correlations are relevant in the present case.

Of particular interest is the behavior of $Z_{\rm eff}$. For all salt
concentrations it is larger than the bare charge and it tends to increase
with increasing salt content. This effect is even more pronounced if the
fits are restricted to macroion separations of $r\ge 1.82\sigma$.  This
finding is in disagreement with the concept of charge renormalization,
where one expects a decrease of the effective charge with increasing salt
concentration. Indeed, cell model calculations using the parameters of
our MD simulations lead to $Z_{\rm eff}\approx 0.8 Z$~\cite{shapran05}.

\begin{figure}[t]
\includegraphics[width=6.cm]{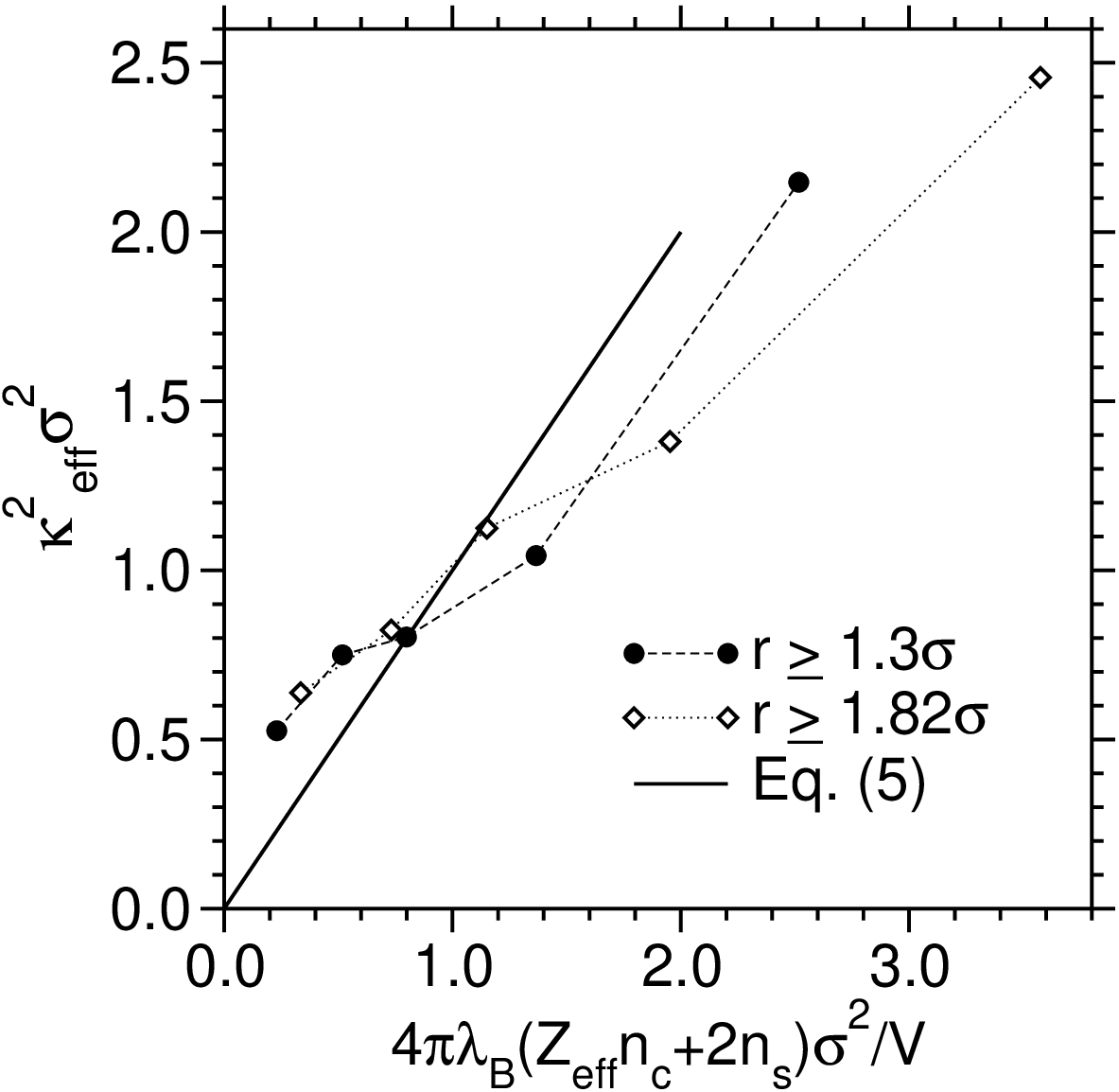}
\caption{
Test of Eq.(\ref{eq:alex}) in the limit $\eta$,$\gamma_0\rightarrow 0$.
\label{fig4}
}
\end{figure}
A failure of the charge renormalization concept in the systems considered
here is not surprising.  For an isolated pair of macroions, there is no
meaningful definition of a WS cell. Hence, it is difficult to define
the volume fraction $\eta$ reasonably. Provided, that the WS cell can
be considered as a sphere around a macroion, we can come up with two
boundaries: The WS cell should not penetrate the Debye layer, thus,
$R\ge r_{\rm ov}^{(1)}/2=\kappa^{-1}+\sigma/2$. In addition, the two
WS cells should not overlap, hence, $R\le r/2$. Note, that the upper
boundary depends on the macroion separation, whereas, according to theory,
$\kappa$ is not a function of $r$. However, our simulation data indicates,
that $\frac{\rm d}{{\rm d}r}\kappa_{\rm eff}$ is slightly different
from zero. Taking into account both limits of $R$, the volume fraction
$\eta=n_{\rm c}(\frac{\sigma}{2R})^3$ should fullfill the inequality
$(1+\frac{2}{\kappa\sigma})^{-1}<\eta^{1/3}<\frac{\sigma}{r}$. With 
$\kappa$ from Eq.~(\ref{kappa}), the lower boundary takes values from
$1.6\cdot10^{-2}$ ($n_{\rm s}=0$) to $9.3\cdot10^{-2}$  ($n_{\rm
s}=1280$). The overall volume fraction of macroions in our simulation box
is given by $\eta'=\frac{\pi\sigma^3}{3V}\approx2.1\cdot10^{-3}$. Hence,
in dilute systems, $\eta'$ cannot be regarded as the relevant volume
fraction for testing Alexander's charge renormalization concept. Moreover,
$\eta'\ll1$ indicates, that Eq.~(\ref{eq:alex}) should hold at least for
small salt concentrations, provided that $\gamma_0$ can be neglected,
i.e., $\kappa_{\rm eff}\approx \kappa$. Although both values, $Z_{\rm
eff}$ and $\kappa_{\rm eff}$, are not found to be in agreement with DH
theory, their relation seems to be compatible with Eq.~(\ref{eq:alex}),
as can be seen from  Fig.~\ref{fig4}. Thus, to a good approximation,
the effective salt concentration matches the actual one, i.e., the Donnan
effect is indeed negligible.

\begin{figure}[t]
\includegraphics[width=6.cm]{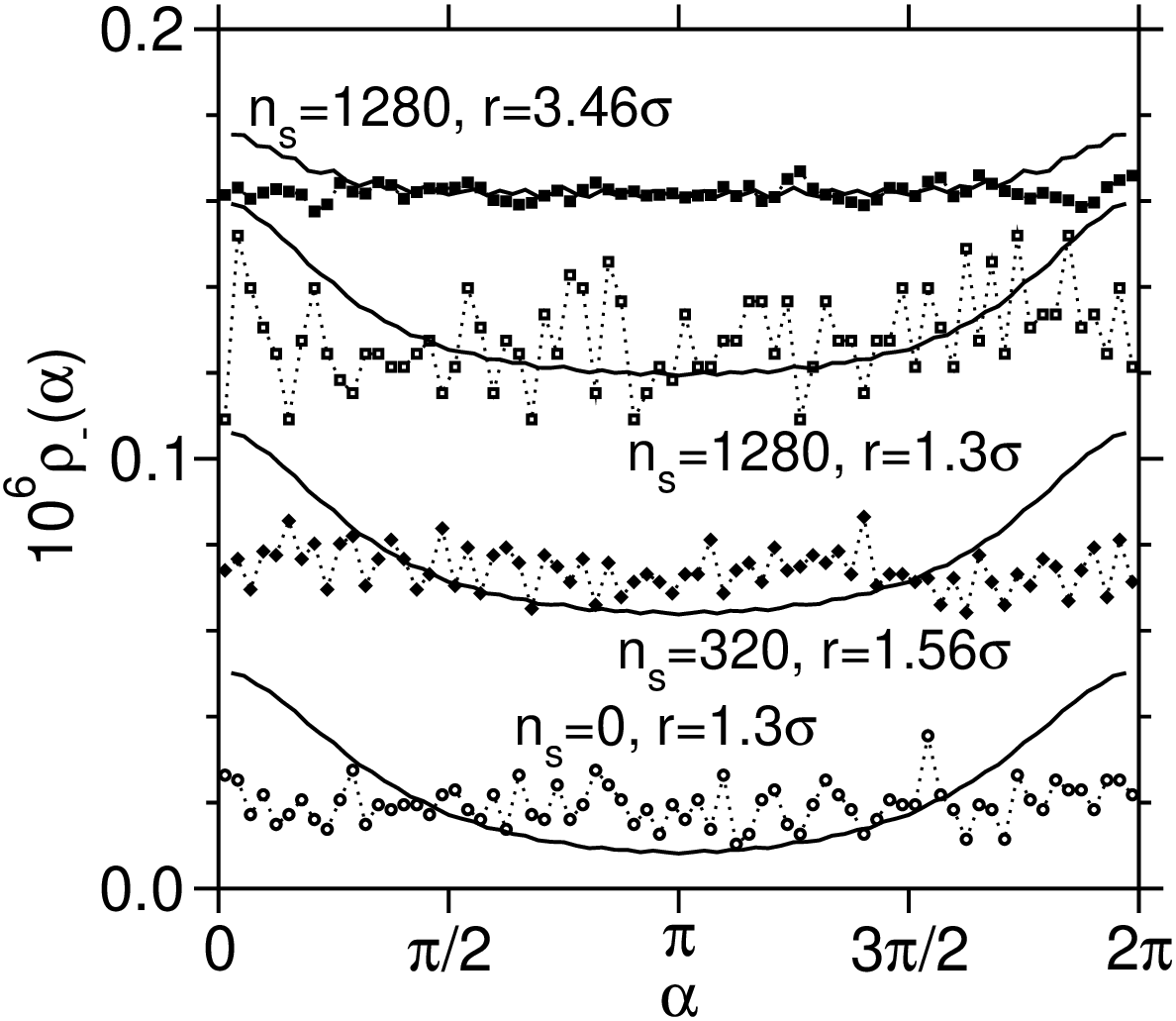}
\caption{
Angular resolved negative charge density distribution around macroion, as
defined by Eq.~(\ref{rhominus.eq}), for the indicated salt concentrations
and macroion separations. Data sets are shifted by $0$, $0.05$,
$0.1$, and $0.15$ (from below). The solid lines are calculated from
the superposition of one--particle charge density distributions from DH
theory (for details see text).
\label{fig5}
}
\end{figure}
So far, we have addressed only the behavior of effective pair forces. In
order to analyze the microionic degrees of freedom, we calculate the
angular resolved negative charge density distribution, $\rho_-(\alpha)$,
which we define as follows: For a given macroion separation, we draw a
sphere of radius $r/2$ around each macroion and project all counterions,
which are located within this fictitious ``WS cell'', onto a plane,
which contains the macroions' centers. Being $n_{\rm ct}(\alpha)$ the
number of counterions at angle $\alpha$, which is taken relative to the
connecting line to the other macroion, $\rho_-(\alpha)$ is given by
\begin{equation}
\label{rhominus.eq}
\rho_-(\alpha)=10^{-6}n_{\rm ct}(\alpha)\left(\frac{\sigma_{\rm mm}}{r}\right)^3.
\end{equation}
Note, that $\rho_-(\alpha)$ is normalized via the volume of the WS sphere,
$\pi r^3/6$.

Fig.~\ref{fig5} shows $\rho_-(\alpha)$ for various combinations of $\kappa
r$. Four different cases are considered: For $n_{\rm s}=0$, we choose $r$
such that $\kappa_{\rm eff} r \ll \kappa r_{\rm ov}^{(3)}$.  For $n_{\rm
s}=320$, we consider the case $\kappa r_{\rm ov}^{(3)}<\kappa_{\rm
eff}r<\kappa r_{\rm ov}^{(2)}$. For $n_{\rm s}=1280$, we take $\kappa
r_{\rm ov}^{(3)}<\kappa_{\rm eff}r<\kappa r_{\rm ov}^{(2)}$ and
$\kappa_{\rm eff}r \gg \kappa r_{\rm ov}^{(1)}$, respectively.

For an isolated pair of macroions, the electric field around a macroion
only exhibits a spherical symmetry in the limit $\kappa r \rightarrow
\infty$. Therefore, one might expect that $\rho_-(\alpha)$ is not
independent of $\alpha$, and thus, it should reveal anisotropies.
However, within the statistics of our data, we find flat distributions
within the ``WS cell''. This holds even for the smallest value of
$\kappa r$, where the Debye layer around a given macroion is strongly
perturbed by the other macroion. The occurrence of isotropic distributions
$\rho_-(\alpha)$ might be due to nonlinearities, which are of course not
accounted for in the DH limit. In order to rationalize this hypothesis,
we checked whether $\rho_-(\alpha)$ can be ``reconstructed'' by a naive
superposition of counterion charge distributions around a single macroion.
To this end, we considered first such single--particle distributions
as obtained from DH theory using the screening parameter $\kappa$ as
given by Eq.~(\ref{kappa}) and the bare charge $Z$ for the charge of the
macroion [note that the charge distribution is just proportional to the
potential given by Eq.~(\ref{effyukawa}) in the linearized DH limit].
Then, we projected the superposition of the latter distributions onto a
cubic lattice with $10^8$ grid points. From this, we finally calculated
$\rho_-(\alpha)$.  The results are included in Fig.~\ref{fig5} as solid
lines. We clearly see that the so calculated $\rho_-(\alpha)$ are
anisotropic, and this indicates that the flat distributions obtained
from the MD simulations might be due to the occurrence of nonlinearities.

\begin{figure}[t]
\includegraphics[width=6.cm]{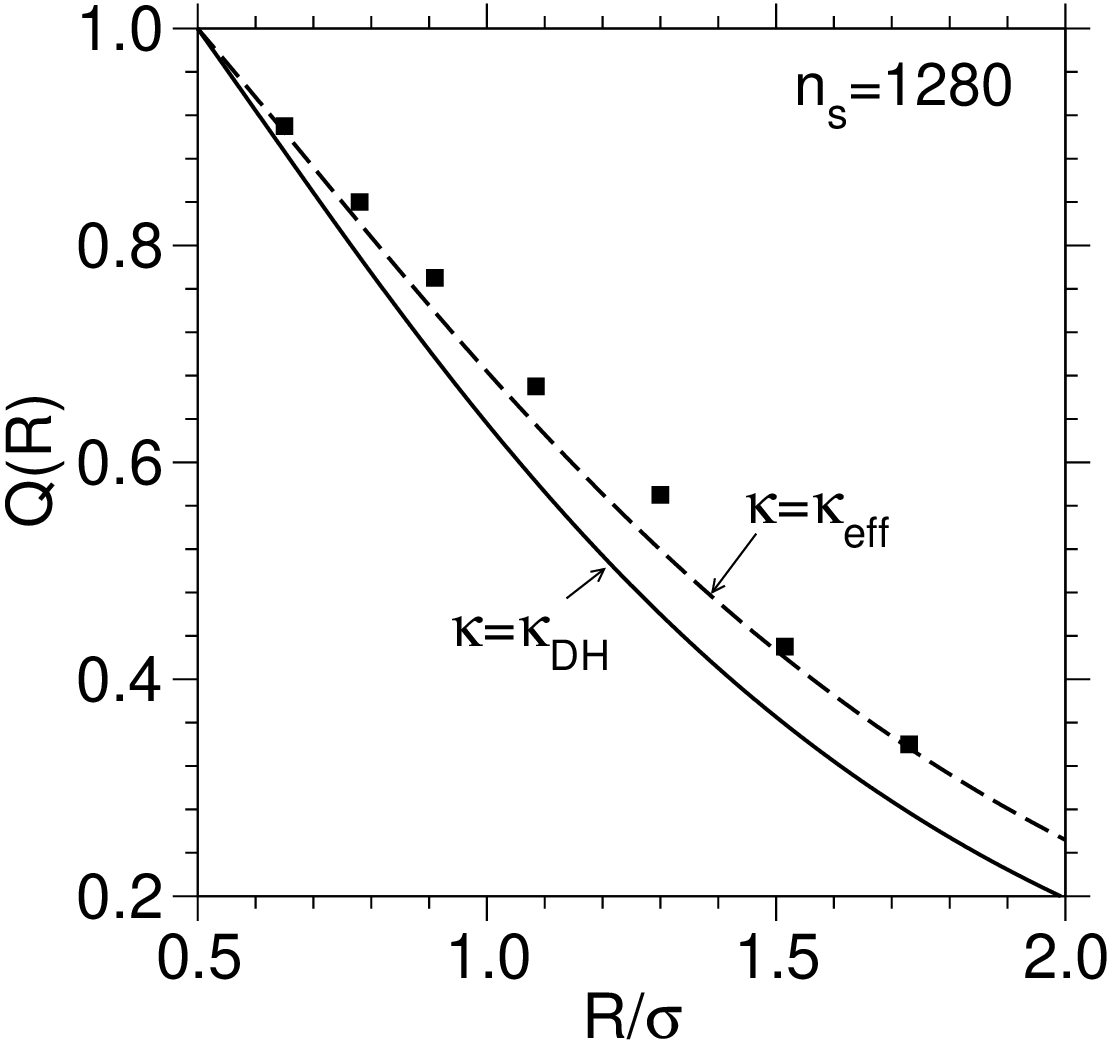}
\caption{
Deviation from charge neutrality within ``WS cell'', measured via $Q(R)$, as
defined in Eq.~(\ref{q.eq}). Data is plotted for $n_{\rm s}=1280$. Results
are compared to Eq.~(\ref{res.eq}), using $\kappa$ from Eq.~(\ref{kappa})
(solid line) and $\kappa_{\rm eff}$ (dashed line).
\label{fig6}
}
\end{figure}
The behavior of $\rho_-(\alpha)$ might also explain why the effective
charge $Z_{\rm eff}$ is higher than the bare charge $Z$, in contrast to
the prediction from the charge renormalization concept. We can infer from
the angular distributions $\rho_-(\alpha)$ that there are less counterions
between the macroions than expected from a naive superposition principle.
This effect might be of entropic origin indicating that the entropy gain
related to isotropic distributions dominates over energetic contributions.
However, energetically unfavoured microion distributions might yield an
additional repulsion between the macroions, and this might explain the
finding that $Z_{\rm eff}$ is larger than the bare charge.

We have already mentioned that the introduction of a ``WS cell'' is not
appropriate for a system of two isolated macroions and thus cell
models that lead to charge renormalization cannot be applied. There
is also another reason why the concept of charge renormalization is
not appropriate in the present case. If we define the boundary of the
(spherical) ``WS cell'' by the sphere of radius $R=r/2$ around a macroion,
this cell is not a neutral object, i.e.~the total charge inside the cell
is nonzero.  This is in contrast to the assumptions of Alexander's cell
model which is not applicable for small macroion separations.

However, it is instructive to study the total charge of the ``WS cell''
for our system. Charge neutrality requires
\begin{equation}
  \label{eqneu}
  Z\left(\frac{\sigma_{\rm m}}{2R}\right)^3+\int_0^{2\pi} 
  {\rm d}\alpha\rho_+(\alpha)
   =\int_0^{2\pi} {\rm d}\alpha\rho_-(\alpha). 
\end{equation}
where $\rho_+(\alpha)$ has an analogous definition as $\rho_-(\alpha)$,
but now the number of counterions at angle $\alpha$ is replaced by the
corresponding number of coions. It follows from Eq.~(\ref{eqneu}) that
\begin{equation}
\label{q.eq}
   Q(R)\equiv 1+\frac{1}{Z}\left(\frac{2R}{\sigma_{\rm m}}\right)^3
   \int_0^{2\pi} {\rm d}\alpha\big[\rho_+(\alpha)-\rho_-(\alpha)\big]
\end{equation} 
should vanish, if the overall charge within the ``WS cell'' is zero.
In Fig.~\ref{fig6}, we show $Q(R)$ for a fixed salt concentration of
$n_{\rm s}=1280$. It is compared to an estimate, which follows from
DH theory: Suppose, the counterion density around a macroion is given
by an expression of the DH form,
\begin{equation}
\label{r.eq}
\rho(r)= \frac{\gamma\exp[-\kappa(r-\frac{\sigma}{2})]}
{(1+\frac{\kappa\sigma}{2})r\lambda_{\rm B}^2},
\end{equation}
where $\gamma$ is a dimensionless normalization constant. Since
$Q(R)=1-\frac{1}{Z}\int_{V(r)}{\rm d}^3r\rho({\bf r})$, the total
charge reads
\begin{eqnarray}
Q(R) &=& 1-\frac{4\pi}{Z}\int_{\sigma/2}^R{\rm d}rr^2\rho(r) \nonumber \\ 
&=& 1 - \frac{4\pi \gamma}{Z\lambda_{\rm B}^2(1+\frac{\kappa\sigma}{2})}
\Bigg[\frac{\sigma}{2\kappa}+\frac{1}{\kappa^2} \nonumber \\ 
&-& \left(\frac{R}{\kappa}+\frac{1}{\kappa^2}\right)
\exp\bigg[-\kappa\sigma\Big(\frac{R}{\sigma}-\frac{1}{2}\Big)\bigg]\Bigg]. 
\end{eqnarray}
The normalization constant introduced in Eq.~(\ref{r.eq})
is determined by the boundary condition $Q(R\rightarrow
\infty)\rightarrow 0$, thus, $\gamma=\frac{Z\lambda_{\rm
B}^2(1+\kappa\sigma/2)}{4\pi(\sigma/2\kappa+1/\kappa^2)}$. Hence,
the charge inside the fictitious WS cell of radius $R$ is given by
the expression
\begin{equation}
\label{res.eq}
Q(R)=\frac{1+ \kappa R}{1+ \frac{\kappa \sigma}{2}}
\exp\Big[-\kappa\sigma\Big(\frac{R}{\sigma}-\frac{1}{2}\Big)\Big].
\end{equation}
Note, that the second boundary condition, $Q(R=\frac{\sigma}{2})=1$,
is intrinsically fulfilled. If we identify the inverse screening
length as $\kappa$ from Eq.~(\ref{kappa}), Eq.~(\ref{res.eq}) slightly
underestimates $Q(R)$. Replacing $\kappa$ by the effective
inverse screening length leads to an almost perfect agreement with
the results of our simulations.

\subsection{Three macroions}

In this section, we consider systems with three macroions in two
different geometries by placing them along a line or at the corners of
an equilateral triangle.

In general, the interaction energy for three particles can be written as
\begin{equation}
V(r)=V_{12}(r_{12})+V_{13}(r_{13})+V_{23}(r_{23})+V_{123}(r_{123}),
\end{equation}
where $V_{ij}(r_{ij})$ is the pair potential between particle $i$ and
$j$. The last term on the right hand side represents the three-body
interactions.

We measure the force on the outermost particle (particle "1") and
define the relative deviation of the electric field with respect to the
expectation for pairwise additivity by
\begin{eqnarray}
\label{delta.eq}
\nonumber \Delta &\equiv& 
-\frac{\frac{\partial}{\partial r}[V(r)-V(r_{12})-V(r_{13})]}
{\frac{\partial}{\partial r}V(r)} \\
&=& \frac{E^{(3)}-E^{(2)}}{E^{(3)}}.
\end{eqnarray} 
Here, $E^{(2)}$ is the superposition of the two-body
interactions calculated in the previous subsection, whereas
$E^{(3)}=-\frac{1}{Z}\frac{\partial}{\partial r}V(r)$ follows from
the force acting on the outermost particle in the three-macroion
configuration.

For the coaxial geometry, the pair contribution is given by
$E^{(2)}=-\frac{1}{Z}\frac{\partial}{\partial r}\sum_{i>1}V(r_{1i})$.
In the configuration of an equilateral triangle with side length $R$,
one has to take into account, that the forces do not act along the same
direction. If we denote the positions of the macroions by $\vec{R}_i$
($i=1,2,3$) the effective force $F(R)$ is given by the total force
$\vec{F}_1$ on particle 1 projected onto the difference vector
$\vec{d}=\vec{R}_1-\frac{1}{3}(\vec{R}_1+\vec{R}_2+\vec{R}_3)$,
\begin{equation}
  F(R) = \vec{F}_1 \cdot \frac{\vec{d}}{|\vec{d}|} \quad . 
\end{equation}
Thus, the two--body contribution in the equilateral
triangle is $E^{(2)}=-\frac{1}{Z}\frac{\partial}{\partial
r}\sum_{i>1}V(r_{1i})\cos(\pi/6)$.

\begin{figure}[t]
\includegraphics[width=6.cm]{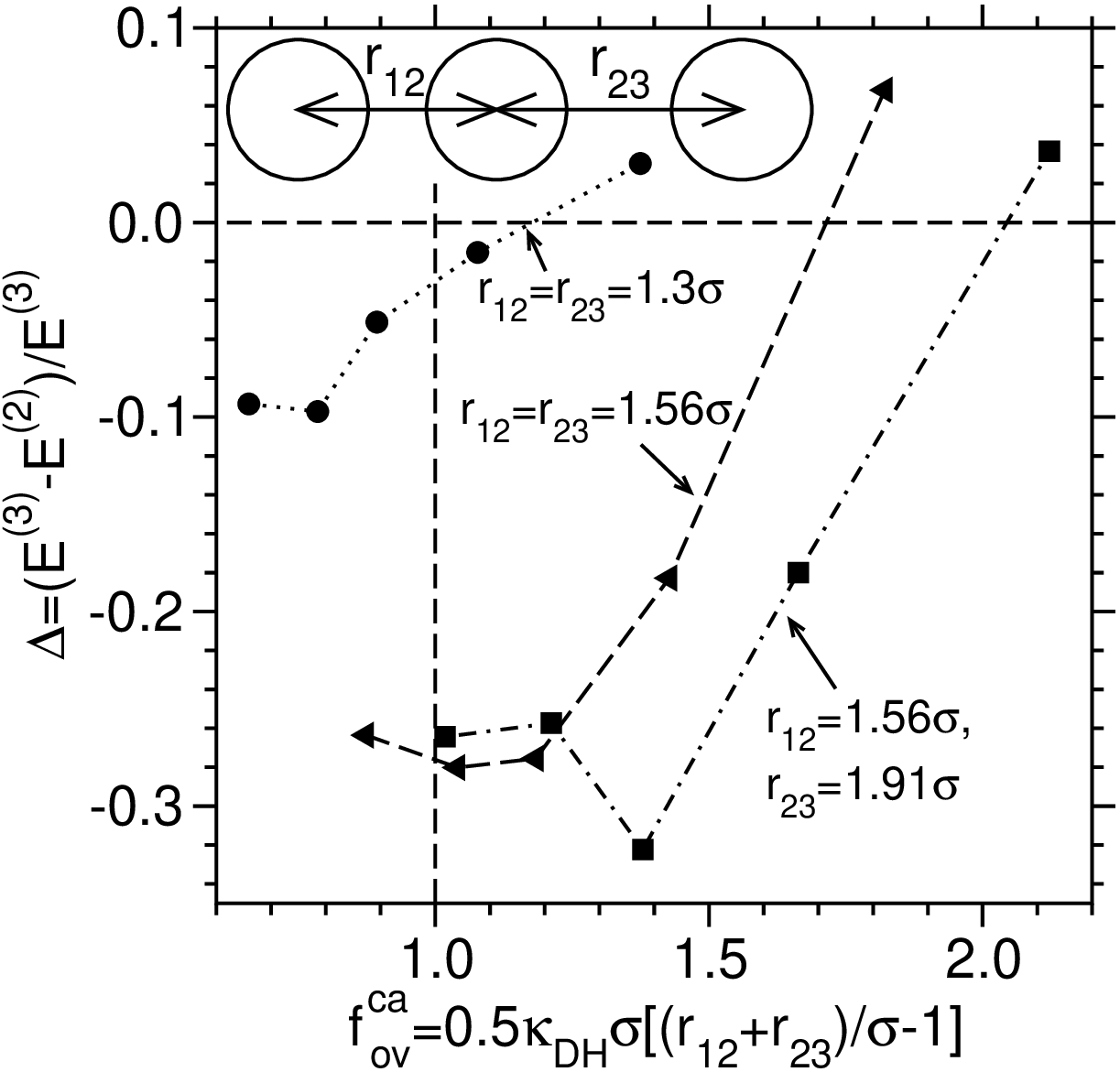}
\caption{
Relative deviation $\Delta$ of electric field around outermost macroion 
in the coaxial geometry as compared to expectation for pairwise additivity
for different distances $r_{12}$ and $r_{23}$ between the particles.
\label{fig7}
}
\end{figure}
Results for the coaxial geometry are displayed in Fig.~\ref{fig7}
where the deviation $\Delta$ from pairwise additivity
is plotted as a function of the parameter $f_{\rm ov}^{\rm
ca}\equiv\frac{\kappa^{(3)}\sigma}{2}[\frac{(r_{12}+r_{23})}{\sigma}-1]$.
The quantity $f_{\rm ov}^{\rm ca}$ describes the overlap between the
Debye layers around the three macroions.  For $f_{\rm ov}^{\rm ca}<1$,
the three Debye layers exhibit an overlap.

One can infer from Fig.~\ref{fig7} that the three-body interaction between
the macroions yields attractive corrections to pairwise additivity. At
small salt concentration, i.e.~if $f_{\rm ov}^{\rm ca}$ is significantly
smaller than one for a given distance between the macroions, three-body
corrections are most pronounced and they are weakly dependent on $f_{\rm
ov}^{\rm ca}$. But if $f_{\rm ov}^{\rm ca}$ reaches values that are of
the order of one, the parameter $\Delta$ increases rapidly and seems
to vanish at high values of $f_{\rm ov}^{\rm ca}$. Thus, three-body
contributions are of importance if there is an overlap between the three
Debye layers. This shows that the range of three-body contributions is
of the order of the Debye length and thus the concept of screening is
also very useful for the discussion of many-body effects.

It is interesting that the three-body terms are much smaller in the
coaxial geometry if the distance between neighboring macroions is close
to contact. This corresponds to the data for $r_{12}=r_{23}=1.3 \sigma$
in Fig.~\ref{fig7}. In this case, the pair interaction probably is the most
important contribution because the three-body force on the outermost
particle is effectively screened out by the particle in the middle.

\begin{figure}[t]
\includegraphics[width=6.cm]{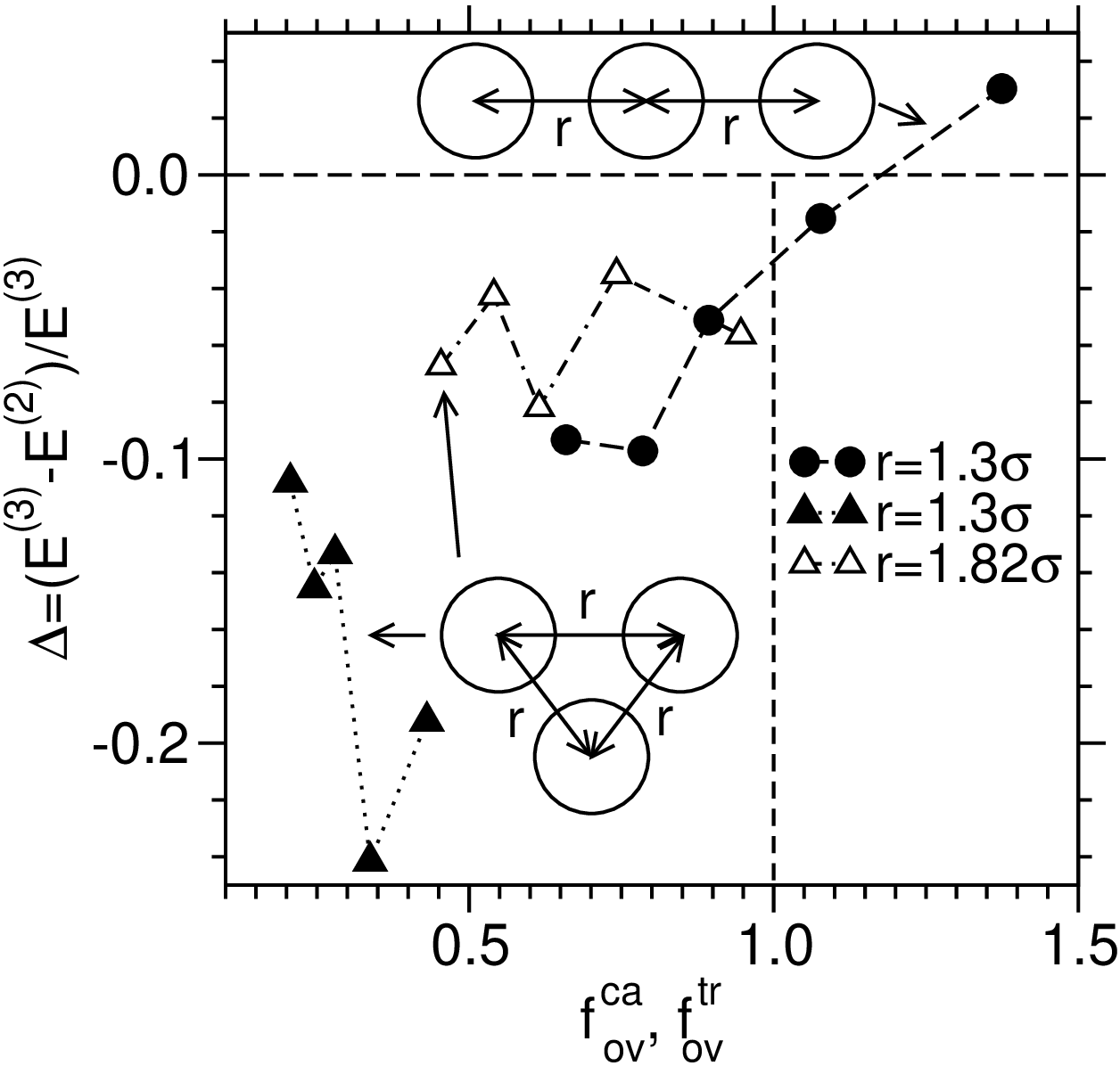}
\caption{
Relative deviation of electric field around macroion as compared to
expectation for pairwise additivity versus overlap factor $f_{\rm ov}$
(see text). Triangular symbols represent triangular setup, circles
represent coaxial geometry.
\label{fig8}
}
\end{figure}
\begin{figure}[t]
\vspace*{0.3cm}
\includegraphics[width=6.cm]{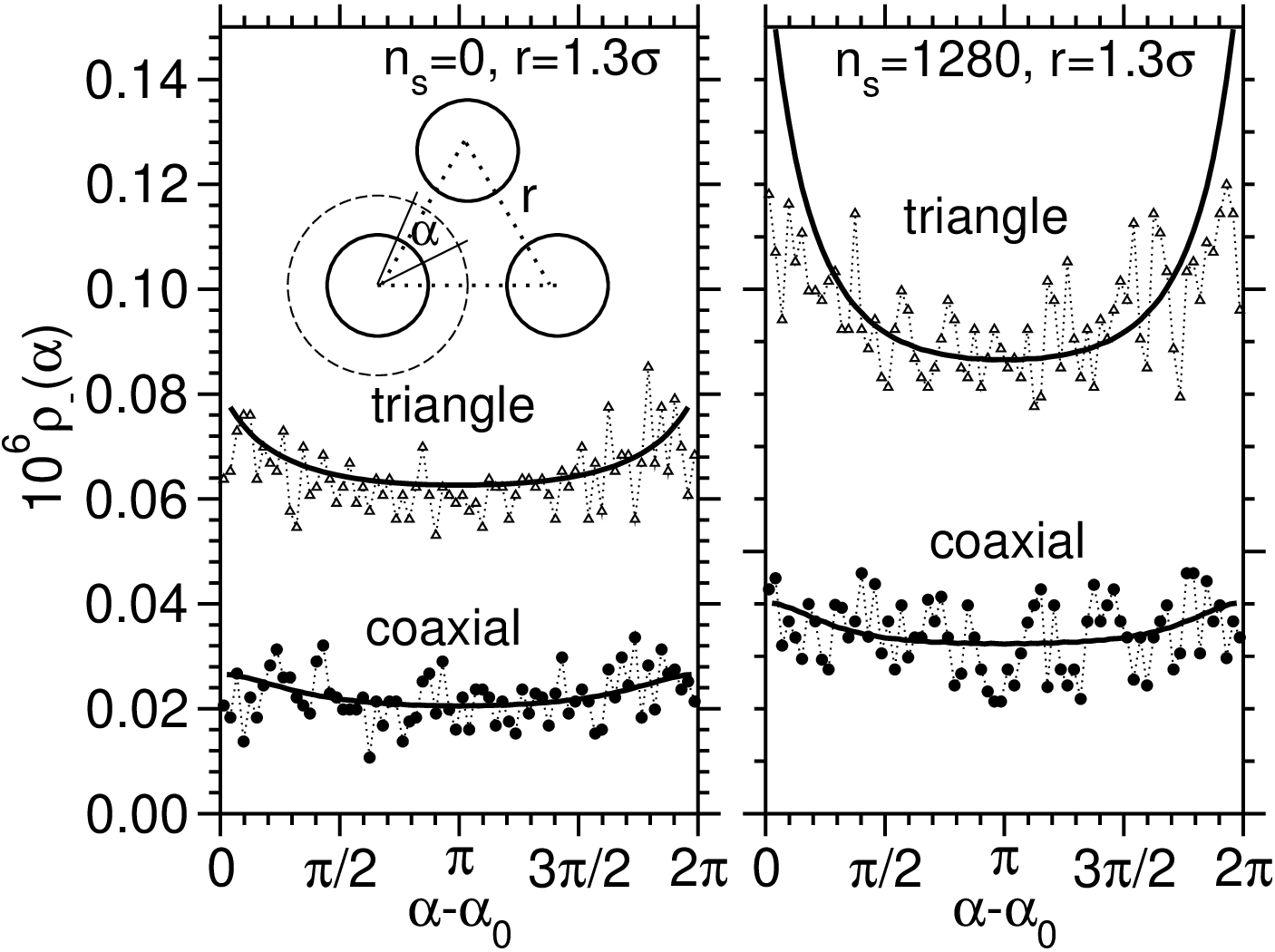}
\caption{
Negative charge density distribution for three-macroion case,
comparing the coaxial geometry to the triangular configuration. The
latter is shifted by $-0.05$. The macroion separation is fixed to
$r=1.3\sigma$. Salt concentration is $n_{\rm s}=0$ (left) and $n_{\rm
s}=1280$ (right), respectively. Solid lines have same meaning as in Fig.~\ref{fig5}.
\label{fig9}
}
\end{figure}
In the case of an equilateral triangle, the condition
for three overlapping Debye layers reads $f_{\rm ov}^{\rm
tr}\equiv\kappa^{(3)}\sigma(\frac{r}{\sqrt{3}\sigma}-\frac{1}{2})<1$.
Using the same values of $\kappa$ and $r$ for both geometries leads
to $f_{\rm ov}^{\rm tr}<f_{\rm ov}^{\rm ca}$, thus, in the triangular
configuration, the Debye layers exhibit a stronger overlap.

In Fig.~\ref{fig8}, three different geometries for the macroion triple
are compared:
\begin{enumerate}
\item an equilateral triangle with side length $r=1.3\sigma$
\item an equilateral triangle with side length $r=1.82\sigma$ 
\item the coaxial geometry with $r=r_{12}=r_{23}=1.3\sigma$. 
\end{enumerate}
The strongest triplet interactions are revealed for case 1. Different from
the coaxial geometry, the magnitude of the  parameter $\Delta$ increases
with decreasing distance $r$ between the particles in the triangle.
This can be easily understood since the interaction between the three
macroions in the triangular geometry is not effectively screened by one
of them but only by the microions in the Debye layers. For case 1 the
magnitude of the parameter $\Delta$ also seems to increase with increasing
$f_{\rm ov}^{\rm tr}$. Up to now, we do not have an explanation for this
behavior.

Comparing case 2 and 3, we see that deviations from pairwise additivity
are similar in both cases, and the overlap parameters $f_{\rm ov}^{\rm
tr}$ and $f_{\rm ov}^{\rm ca}$ have comparable values.  A similar
feature has been found in a numerical solution of the nonlinear PB
equation by Russ {\it et al.}~\cite{russ}.  These authors report that
the three--body potential is independent of geometry if the sum over
the distances between neighboring particles is constant.

We would like to point out that there is always a trivial contribution
to the many--body potential, which stems from an increased microion
concentration associated with the addition of macroions.  Thus, for a
fixed volume, the effective screening length of the system is decreased. A
comparison between the measured three--body term, $E^{(3)}$, and the
two--body contribution, $E^{(2)}$, taken from the pure two-macroion
case should therefore in general yield a non--pairwise additivity. From
that point of view, DH theory already predicts a correction to pairwise
additivity.

Similar to the previous subsection, we calculate the angular
resolved charge density distribution around the outermost macroion [see
Eq.~(\ref{rhominus.eq})]. In order to account for the differences between
coaxial and triangular geometry, we introduce an angle $\alpha_0$, such
that the system is symmetric around $\alpha=\alpha_0$. Thus, we have
to choose $\alpha_0=0$ for the coaxial geometry and $\alpha_0=\pi/6$
for the equilateral triangle.

As Fig.~\ref{fig9} shows, the charge distribution for the three macroion
case is not isotropic.  Compared to the case of two macroions,
the internal energy of the system is now larger, and, therefore,
entropic contributions might be less important. In accordance with
our previous consideration, namely that an energetically unfavored
microion distribution might lead to an additional repulsion between
the macroions, we might conclude here that the onset of anisotropy is
correlated with non--pairwise additivity. The choice of the same value of $r$
leads to a stronger anisotropy for the triangle as compared to the coaxial
geometry, which is consistent with the behavior of the parameter $\Delta$
(see Fig.~\ref{fig8}). As done in the previous section,
we compare our charge distribution to the one which follows from naive
superposition of the DH distributions. Surprisingly, for the three
macroion configurations, this superposition seems to work very well.

\section{Concluding remarks}

We performed classical MD simulations in order to
investigate effective interactions between isolated pairs and
triples of charged macroions in the framework of the primitive model. 

On the pair level, these interactions are surprisingly well described by
the DH limit of the PB equation.  In particular there is no evidence for
charge renormalization as predicted by cell models. These models would
predict an effective charge which is considerably smaller than the bare
charge of the macroion~\cite{alex,trizac,shapran05}.  This finding is not
due to finite size effects in the simulation which might emerge if the
Debye length exceeds the size of the simulation box.  It rather follows
from the fact, that the cell model must not be applied to systems of
isolated macroions.  This means that the concept of charge renormalization might
be relevant for bulk systems, but, in the case of systems of isolated macroions, 
simulations should be compared to direct solutions of the nonlinear PB equation.

In this work, we have studied systems with small salt concentrations
of the order of a few $\mu$Mol, and we have considered configurations
for which the Debye layers of the different macroions exhibit a strong
overlap.  An interesting result of our simulations is the occurrence
of repulsive corrections to DH theory: For the macroion pair, we find
effective macroion charges that are slightly higher than their bare
charge. Similar results have been reported in previous work, e.g., in
an ab initio density functional theory approach \cite{tehver}, where the
ratio between effective and bare charge was found to be between $1.06$
and $1.38$, depending on salt concentration and the value of the bare
charge.  An ``repulsive correction to DH theory'' is also indicated by
the isotropic charge distribution around the macroion in the case of the
macroion pair. Such an isotropic distribution is not expected from a naive
superposition of one--particle density distributions as obtained from DH
theory. Hence, there seem to be less counterions between the macroions
than expected from DH theory which can be related to an increase of the
effective charge.  The microscopic origin of this effect is not clear,
but it might be of entropic origin. 

In agreement with previous analytical \cite{denton04}, numerical
\cite{lowen,russ,wu} and experimental studies\cite{brunner,dobnikar}
of systems with three isolated macroions, we find that corrections to
non--pairwise additivity (and thus the three--body terms in the effective
potential) are attractive. The strength of these attractive contributions
is strongly correlated with the overlap of all three Debye layers. This
shows that the concept of a screening length is also very useful to
quantify the effect of three--body interactions.  Different from the
case of two macroions, the charge distribution in the three--macroion
case is anisotropic. In this case, the simple superposition of three
one--particle density distributions from DH theory yields a rather good
description of the charge distribution in the three--macroion case. This
finding seems to agree with a recent numerical solution of the PB equation
for three isolated macroions \cite{russ}.

In further simulation studies, we will investigate interactions between more
than three particles to understand the crossover to bulk effective
interactions.  In the latter case, the concept of charge renormalization
seems to work very well.  Our present simulations suggest that many--body
interactions in bulk systems yield renormalized charges that can be
much smaller than the bare charges of charged colloidal particles. A
profound understanding of these issues might also provide new insight
into electrophoresis experiments \cite{garbow04,medebach04}.

\begin{acknowledgments}
We are grateful to Thomas Palberg for many helpful discussions.
Financial support of the DFG (SFB 625) and the MWFZ Mainz
are gratefully acknowledged.  Two of the authors (J.H. and A.C.) acknowledge
the support through the Emmy Noether program of the DFG, Grant No. HO 2231/2.
Computing time on the JUMP at the NIC J\"ulich is gratefully
acknowledged.
\end{acknowledgments}

\end{document}